\documentclass[preprint,preprintnumbers,amsmath,amssymb,12pt,floatfix,epsfig,nofootinbib,superscriptaddress]{revtex4}

\newcommand{\stkout}[1]{\ifmmode\text{\sout{\ensuremath{#1}}}\else\sout{#1}\fi}

\usepackage{amsmath} 
\usepackage{bm}      
\usepackage[usenames, dvipsnames]{color}
\usepackage{graphicx}
\usepackage{amssymb}
\usepackage[normalem]{ulem}
\renewcommand\sout{\bgroup\color[rgb]{1,0,0} \ULdepth=-.5ex \ULset}

\begin{document}

\title{Nuclear Pairing Energy vs Mean Field Energy: Do They Talk To Each Other For Searching The Energy Minimum?}
\author{Myeong-Hwan Mun \footnote{aa3101@gmail.com}} 
\affiliation{Department of Physics and Origin of Matter and Evolution of Galaxies (OMEG) Institute, Soongsil University, Seoul 06978, Korea}

\author{Eunja Ha \footnote{ejaha@hanyang.ac.kr}}
\affiliation{Department of Physics and Research Institute for Natural Science, Hanyang University, Seoul 04763, Korea}

\author{Myung-Ki Cheoun \footnote{Contact author: cheoun@ssu.ac.kr}}
\affiliation{Department of Physics and Origin of Matter and Evolution of Galaxies (OMEG) Institute, Soongsil University, Seoul 06978, Korea}

\author{Yusuke Tanimura \footnote{tanimura@ssu.ac.kr}}
\affiliation{Department of Physics and Origin of Matter and Evolution of Galaxies (OMEG) Institute, Soongsil University, Seoul 06978, Korea}

\author{Hiroyuki Sagawa \footnote{sagawa@ribf.riken.jp}}
\address{RIKEN, Nishina Center for Accelerator-Based Science, Wako 351-0198, Japan and Center for Mathematics and Physics, University of Aizu, Aizu-Wakamatsu, Fukushima 965-8560, Japan}
\author{Gianluca Col\`o \footnote{colo@mi.infn.it}}
\address{Dipartimento di Fisica, Universit\`a degli Studi di Milano, via Celoria 16, 20133 Milano, Italy}
\address{INFN, Sezione di Milano, via Celoria 16, 20133 Milano, Italy}

\date{\today}

\begin{abstract}
We study the evolution of the total binding energy  (TBE) and pairing energy of Pb, Hg and Ar isotopes, as a function of the nuclear deformation. As for the nuclear model, we exploit a deformed relativistic Hartree-Bogoliubov theory in the continuum (DRHBc), and a deformed Skyrme Hartree-Fock plus BCS model. It is found that the dependence of pairing energy on the deformation is strongly correlated to that of the mean field energy, which is obtained by subtracting the pairing energy from the TBE; in other words, the energy minimum characterized by a large negative mean field energy has a smaller negative pairing energy or, equivalently, a smaller positive pairing gap, while 
a stronger pairing energy is found in the region away from the minimum of the total energy. Consequently, the two energies show an anti-symmetric feature in their deformation dependence,  although the energy scales are very different.  Moreover, since the pairing energy has a negative sign with respect to to the pairing gap, the evolution of mean field energy  follows closely that of the 
pairing gap. This implies that the pairing energy (or pairing gap) and the mean field energy talk to each other and work together along the potential energy curve to determine the energy minimum and/or the local minimum.

\end{abstract}

\maketitle

\section{Introduction}
Recent developments of microscopic nuclear models enable us to predict many intriguing properties such as 
deformation and shape coexistence, nuclear bubble structures, odd-even staggering of binding energies and radii and, last but not least, the location of the drip lines that marks the limit of the existence
of nuclei, in a wide region of the nuclear chart. In particular, the shape coexistence in heavy and superheavy nuclei \cite{Poves2016,Gade2016,Heyde2011,Nacher2004,Andreyev2000,Ojala2022} reveals important information on nuclear shapes or nuclear shape transitions, deeply associated with nuclear deformation and the rotational band structure. The nuclear deformation is also intimately related to the nuclear surface contribution to the symmetry energy in finite nuclei \cite{Pawel2003,Mun2024}.

On the other hand, pairing of fermions or pairing correlations play a crucial role in superconducting solids \cite{Bardeen1957}, and also constitute an important complement to the nuclear shell structure \cite{Bohr1958}. {Most frequently, the pairing phenomena are discussed in} the so-called Bardeen-Cooper-Schrieffer (BCS) approximation \cite{Ring1980,Nilsson1995}, which was introduced in the original paper of Ref. \cite{Bardeen1957} {for the metal superconductivity}. With the recent development of non-relativisitc or covariant  Density Functional Theory (DFT) for the nuclear many-body problem, various prescriptions for the pairing energy functional have been suggested. They basically consist in defining a small band of pairing-active states (pairing window) and some effective strength depending on the system size, {\it i.e.} neutron and proton numbers; local two-body pairing forces with more or less sophisticated treatments of the isoscalar and isovector neutron-proton pairing correlations can be defined \cite{Yama2012,Bai2013,Sagawa2016}.

In this work, we study the evolution of total binding energies (TBEs) in  Pb, Hg and Ar isotopes as a function of the nuclear deformation $\beta_2$,  and compare it with the evolution of pairing energies and of the binding energies obtained by subtracting the pairing energy, {\it i.e.} the mean field  energies. 
As for the nuclear model, we exploit the deformed relativistic Hartree-Bogoliubov approach in the continuum (DRHBc) approach \cite{Kaiyuan2020,Kaiyuan2021,Cong2021,Kaiyuan2022,Peng2024}. The  evolution of those energies with respect to the deformation is also tested by deformed Skyrme Hartree-Fock-BCS (DSHF+BCS) calculations. The aim of this study is to understand how the pairing energy will correlate with the mean field  energy. In other words, we would like to find out the cross-talk between the deformation minimum and the pairing energy maximum quantitatively  in the mean field calculations. The existence of the pairing correlations always bring more binding energy to a nucleus, but for doubly-magic nuclei which are more stable than neighbours, pairing correlations do not contribute to the total energy of ground state. Therefore, it would be an intriguing task how the pairing correlations are related to the mean field energy for other nuclei.

This paper is organized as follows. In Sec. II, we briefly summarize the basic formalism used in the present calculation. Detailed results for the mean field energies, the pairing energies, and the pairing gaps in terms of the deformation for Pb, Hg and Ar isotopes are provided in Sec. III. Finally, summary and conclusion are presented in Sec. IV.

\section{FORMALISM}
In order to properly carry out a quantitative discussion along the lines we mentioned, we need a well-refined nuclear model which incorporates the deformation, the pairing correlations and the continuum through a  microscopic approach, and which explains the whole nuclear masses covering nuclei near drip-lines.
To properly describe the cross-talk between mean-field and pairing energies, we need a consistent, state-of-the-art model. Pairing and deformation should be treated self-consistently, and the present approach also incorporates continuum coupling, so that our analysis can be made throughout the mass table up to the drip lines.
The relativistic mean field model has the advantage of describing the evolution of spin-orbit splitting in a wide region of the 
mass table without introducing any free parameters \cite{Walecka1974,Boguta1977}. We compare some of the DRHBc results with those of the non-relativistic Skyrme energy density functionals (EDFs) to check the role of exchange terms in EDFs, since they are not taken into account in the DRHBc model.

The DRHBc model was developed for deformed halo nuclei in Refs. \cite{Zhou2010,Lulu2012},  and recently extended by exploiting the  point-coupling EDFs \cite{Kaiyuan2020}. 
This approach has been proven to be capable of giving a good description of the nuclear masses with highly predictive power \cite{Kaiyuan2021,Cong2021}, and successfully applied to  nuclei in a wide region of the mass table \cite{Cong2019,Sun2018,Sun2020,In2021,Yang2021,SunPRC1,SunPRC2,Sun2021,Mun2023}, 
following the previous relativistic continuum Hartree-Bogoliubov (RCHB) approach formulated in coordinate space \cite{Meng1996,Meng1998},  
and explicitly including the deformation in a Dirac Woods-Saxon basis by Legendre expansion \cite{Zhou2003}. 

In this work, we adopt the DRHBc approach with a density-dependent zero-range pairing interaction, which was succinctly summarized in Refs. \cite{Kaiyuan2020,Lulu2012}. We start from the following relativistic mean field (RMF) Lagrangian \cite{Kaiyuan2020}
\begin{eqnarray}
{\cal L} &=& {\bar \psi} ( i \gamma_{\mu} \partial^{\mu} - M) \psi - {1 \over 2} \alpha_S ({\bar \psi} \psi)({\bar \psi} \psi) - {1 \over 2} \alpha_V  ({\bar \psi} \gamma_{\mu} \psi)({\bar \psi} \gamma^{\mu} \psi) - {1 \over 2} \alpha_{TV}  ({\bar \psi} {\vec \tau} \gamma_{\mu} \psi)({\bar \psi} {\vec \tau} \gamma^{\mu} \psi)  \nonumber \\
& & - {1 \over 2} \alpha_{TS}  ({\bar \psi} {\vec \tau} \psi)({\bar \psi} {\vec \tau} \psi) 
- {1 \over 3} \beta_S {({\bar \psi} \psi)}^3 - {1 \over 4} \gamma_S {({\bar \psi} \psi)}^4 
- {1 \over 4} \gamma_V  {[({\bar \psi} \gamma_{\mu} \psi)({\bar \psi} \gamma^{\mu} \psi)]}^2  \nonumber \\
& & - {1 \over 2} \delta_S \partial_{\nu} ({\bar \psi} \psi) \partial^{\nu} ({\bar \psi} \psi) - {1 \over 2} \delta_V \partial_{\nu} ({\bar \psi} \gamma_{\mu} \psi) \partial^{\nu} ({\bar \psi} \gamma^{\mu} \psi) 
- {1 \over 2} \delta_{TV}  \partial_{\nu} ({\bar \psi} {\vec \tau} \gamma_{\mu} \psi) \partial^{\nu} ({\bar \psi} {\vec \tau} \gamma^{\mu} \psi)  \nonumber \\
& & - {1 \over 2} \delta_{TS} \partial_{\nu} ({\bar \psi} {\vec \tau} \psi) \partial^{\nu} ({\bar \psi} {\vec \tau} \psi)
- { 1\over 4} F^{\mu \nu} F_{\mu \nu} - e {\bar \psi} \gamma^{\mu} {  {1 -\tau_3} \over 2}  A_{\mu} \psi ~,
\end{eqnarray}
where  $M$ is the nucleon mass, $e$ is the proton charge, and $A_\mu$ and $F_{\mu\nu}$ are the four-vector potential and field tensor of the electromagnetic field, respectively. The coupling constant $\alpha_i$ for four-fermion terms is specified by superscripts $(i=S, V$ and $T)$ which stand for the scalar, vector, and isovector (IV) channels, respectively.  The higher-order terms are specified
by $\beta_i$ and $\gamma_i$,  while $\delta_i$ refers to gradient terms. 
The IV-scalar channels, $\alpha_{TS}$ and $\delta_{TS}$, are neglected in the DRHBc approach \cite{Kaiyuan2020}. 
The EDF is derived in terms of nucleon scalar density, $\rho_S ({\bm r}) = \mathit{\sum_{i=1}^A} {\bar \psi}_i ({\bm r}) {\psi}_i (\bm r)$, and isoscalar $(IS)$ and $IV$ currents, $j_{\mu} ({\bm r}) = \mathit{\sum_{i=1}^A}{\bar \psi}_i ({\bm r}) \gamma_{\mu} \psi_i ({\bm r})$ and ${\vec j}_{\mu} ({\bm r}) =\mathit{\sum_{i=1}^A} {\bar \psi}_i({\bm r}) {\vec \tau}\gamma_{\mu} \psi_i ({\bm r})$, as follows \cite{Meng1998,Niksic2014}
\begin{eqnarray}
E_{RMF} [ \psi, {\bar \psi}, A_{\mu}] &=& \sum_{i=1}^{A} \int d^3 {\bm r} \psi_i^+ ({\bm \alpha} \cdot {\bm p} + \beta M) \psi_i - { 1 \over 2} {(\nabla \cdot {\bm  A})}^2 + {1 \over 2} e \int d^3 {\bm r} j_p^{\mu} A_{\mu} \nonumber \\
& & + \int d^3 {\bm r} [ { 1 \over 2} (\alpha_S \rho_S^2 + \alpha_V j^{\mu} j_{\mu} + \alpha_{TV} {\vec j}^{\mu} \cdot {\vec j}_{\mu}) + { 1 \over 3} \beta_S \rho_S^3 + { 1 \over 4} \gamma_S \rho_s^4 \nonumber \\
& &  +  {1 \over 4} \gamma_V {(j^{\mu} j_{\mu})}^2 + { 1 \over 2} \delta_S \rho_S \square \rho_S + { 1 \over 2} \delta_V  j^{\mu} \square j_{\mu} + { 1 \over 2} \delta_{TV}  {\vec j}^{\mu} \cdot \square {\vec j}_{\mu}  ]  ~. 
\end{eqnarray}
The present calculations are carried out by solving the relativistic Hartree Bogoliubov (RHB) equation with the density functional PC-PK1 \cite{Zhao2010}. The RHB equation reads \cite{Meng1998} 
\begin{equation} \label{eq:hfbeq}
\int d^3 {\bm r'} \left(  \begin{array}{cc} h_D ({\bm r},{\bm r'})- \lambda_\tau &
\Delta ({\bm r},{\bm r'}) \\
 - \Delta^{*} ({\bm r},{\bm r'})& - h^{*}_D ({\bm r},{\bm r'})+ \lambda_\tau 
  \end{array}\right)
\left( \begin{array}{c}
U_{k} {(\bm r)} \\ V_{k} {(\bm r)} \end{array}\right)
 =
 E_{k}
\left( \begin{array}{c} U_{k} {(\bm r)} \\
V_{k} {(\bm r)} \end{array}\right),
\end{equation}
where $\lambda_\tau$ is the Fermi energy ($\tau = \mathrm{n}/\mathrm{p}$) for neutrons or protons. $E_k$ and $(U_k,V_k)^{T}$ are the quasiparticle energy and quasiparticle wave function. In coordinate space, the Dirac Hamiltonian $h_D$ is given by
\begin{equation} \label{eq:hd}
h_D ({\bm r},{\bm r'})= [ {\bm {\alpha}} \cdot {\bm p}  +  V(\bm{r})  +  \beta (M + S({\bm r})) ] \delta ({\bm r},{\bm r'}), 
\end{equation}
where $M$ is the nucleon mass, and $V(\bm{r})$ and $S(\bm{r})$ are the vector and scalar potentials, respectively, given by
\begin{eqnarray}
S( {\bm r}) &=& \alpha_S \rho_S + \beta_S \rho_S^2 + \gamma_S \rho_S^3 + \delta_S \Delta \rho_S ~, \\ \nonumber
V( {\bm r}) &=& \alpha_V \rho_V + \gamma_V \rho_V^3 + \delta_V \Delta \rho_V + e A^0 + \alpha_{TV} \tau_3 \rho_3 + \delta_{TV} \tau_3 \Delta \rho_3 ~,
\end{eqnarray}
with the following densities represented in terms of the quasiparticle wave functions as follows
\begin{equation}
\rho_S ( {\bm r} ) = \sum_{k>0} {\bar V}_k ( {\bm r}) V_k ({\bm r})~,~
\rho_V ( {\bm r} ) = \sum_{k>0} {\bar V}_k ( {\bm r}) \gamma_0 V_k ({\bm r})~,~
\rho_3 ( {\bm r} ) = \sum_{k>0} {\bar V}_k ( {\bm r}) \gamma_0 \tau_3  V_k ({\bm r}).
\end{equation}
The paring potential $\Delta ({\bm r},{\bm r'})$ is given by the following gap equation in terms of the pairing tensor {$\kappa ({\bm r}, {\bm r}^{'})$} \cite{Lulu2012}
\begin{equation} \label{eq:gap}
\mathit{\Delta}({\bm r}, {\bm r}^{'}) = V^{pp} ({\bm r}, {\bm r}^{'}) \kappa ({\bm r}, {\bm r}^{'})~,~
\kappa ({\bm r}, {\bm r}^{'}) = \sum_{k>0} V_k^* ( {\bm r} ) U_k^T ({\bm r'})~,
\end{equation}
where we did not write explicitly the sum on the spin degree of freedom and the upper and lower components of Dirac spinors. The DRHBc approach uses a density-dependent zero-range force
\begin{equation}\label{eq:pairing}
\mathit{V}^{pp} ({\bm r}, {\bm r}^{'}) =  \frac{V_0}{2}( 1 - P_{\sigma}) \delta ( {\bm r} - {\bm r}^{'}) ( 1 -  { \rho (\bm r) \over \rho_{\mathrm{sat}}} )~,
\end{equation}
where $\rho_{\mathrm{sat}}$ is the nuclear saturation density, $V_0$ is the pairing strength, and $(1-P_{\sigma}) / 2$ is the projector for the spin-singlet, $S = 0$ component in the pairing channel. 
Then the pairing potential and pairing tensor of Eq. (\ref{eq:gap}) become local quantities \cite{Lulu2012},
\begin{equation}
\mathit{\Delta}({\bm r}) = V_0 ( 1 - \rho ( {\bm r} )/\rho_{sat}) \kappa ({\bm r}) ~,~  \kappa ({\bm r}) = \sum_{k >0} V_k^{+} ({\bm r}) U_k ({\bm r})
\end{equation}
because of the zero-range and spin-singlet properties of the pairing force in Eq. (\ref{eq:pairing}). For the non-local and spin-triplet (isoscalar) interactions, one needs a more careful treatment of the pairing potential and pairing tensor \cite{Meng1998}. 

Finally, the total energy of a nucleus including the pairing energy and the center of mass (c.m.) correction is defined by 
\begin{eqnarray}\label{eq:EDF}
\mathit{E}_{\rm tot + cm} &=& \mathit{E}_{\mathrm{EDF}}+ \mathit{E}_{\mathrm{pair}} + \mathit{E}_{\mathrm{cm}},
\end{eqnarray}
where $\mathit{E}_{\mathrm{EDF}}$ can be expressed with the quasiparticle energies $E_k$ and wave functions of Eq. (\ref{eq:hfbeq})  as
\begin{eqnarray}\label{eq:EDF2}
 \mathit{E}_{\mathrm{EDF}}&=&
{\sum_{k > 0} \int d^3 {\bm r} (\lambda_{\tau} - E_k) V_k^+ ( {\bm r}) V_k ({\bm r})}  \nonumber \\
&& - \int \mathrm{d}^3{\bm r} \left( \frac{1}{2} \alpha_S \rho_S^2 
+ \frac{1}{2} \alpha_V \rho_V^2 
+ \frac{1}{2} \alpha_{TV} \rho_{TV}^2 
+ \frac{2}{3} \beta_S \rho_S^3 
+ \frac{3}{4} \gamma_S \rho_S^4 
+ \frac{3}{4} \gamma_V \rho_V^4 \right. \nonumber \\
&& \left. + \frac{1}{2} \delta_S \rho_S \Delta \rho_S 
+ \frac{1}{2} \delta_V \rho_V \Delta \rho_V 
+ \frac{1}{2} \delta_{TV} \rho_{TV} \Delta \rho_{TV} 
+ \frac{1}{2} \rho_p e A^0 \right)
- 2\mathit{E}_{\mathrm{pair}}.
\end{eqnarray}
We note that this EDF can be derived in the general HFB case, as shown in Appendix A.  

In the DRHBc approach \cite{Cong2022}, the pairing energy $E_{\mathrm{pair}}$, with the zero-range pairing force, is given by the pairing potential and pairing tensor as follows \cite{Cong2022}
\begin{equation}\label{eq:Penergy}
E_{\mathrm{pair}} = -\frac{1}{2}\int\mathrm{d}^3{\bm r} \kappa(\bm{r})\Delta(\bm{r}) .
\end{equation}

For the pairing strength, we use $V_0$ = -- 325.0 MeV fm${^3}$. The saturation density is adopted as $\rho_{sat}$ = 0.152 fm$^{-3}$ together with a pairing window of 100 MeV. The energy cutoff $E_{cut}^+ =$ 300 MeV,  and the angular momentum cutoff $J_{max} = (23/2) \hbar $, are taken for the Dirac Woods- Saxon basis. The above numerical details are the same as those suggested in Refs. \cite{Kaiyuan2020,Kaiyuan2021} for the DRHBc mass table calculations. For the present calculation, the Legendre expansion truncation is chosen as ${\lambda}_{max}$ = 8 \cite{Kaiyuan2020,Kaiyuan2021}.

Empirical pairing gaps of Pb isotopes were shown to be properly reproduced with the energy cutoff, the maximum angular momentum, and the Legendre expansion truncation obtained from the convergence check of total energies as shown in Fig. 5(b) in Ref. \cite{Kaiyuan2020}.

The present zero-range scheme for the pairing force is better than the simple constant gap approximation, but it has still the pairing window problem in the pairing tensor as discussed in Refs. \cite{Dobaczewski1996,Tian2009},  because it needs an arbitrary energy cut off parameters for neutron-rich nuclei. In spherical nuclei, the neutron pairing gaps are well reproduced by the pairing window denoted as $E_{cut}^{q.p}$ = 100 MeV. 


\section{RESULTS AND DISCUSSIONS}
\subsection{Pairing energy and mean field energy}

\begin{figure}
\centering
\includegraphics[width=0.90\linewidth]{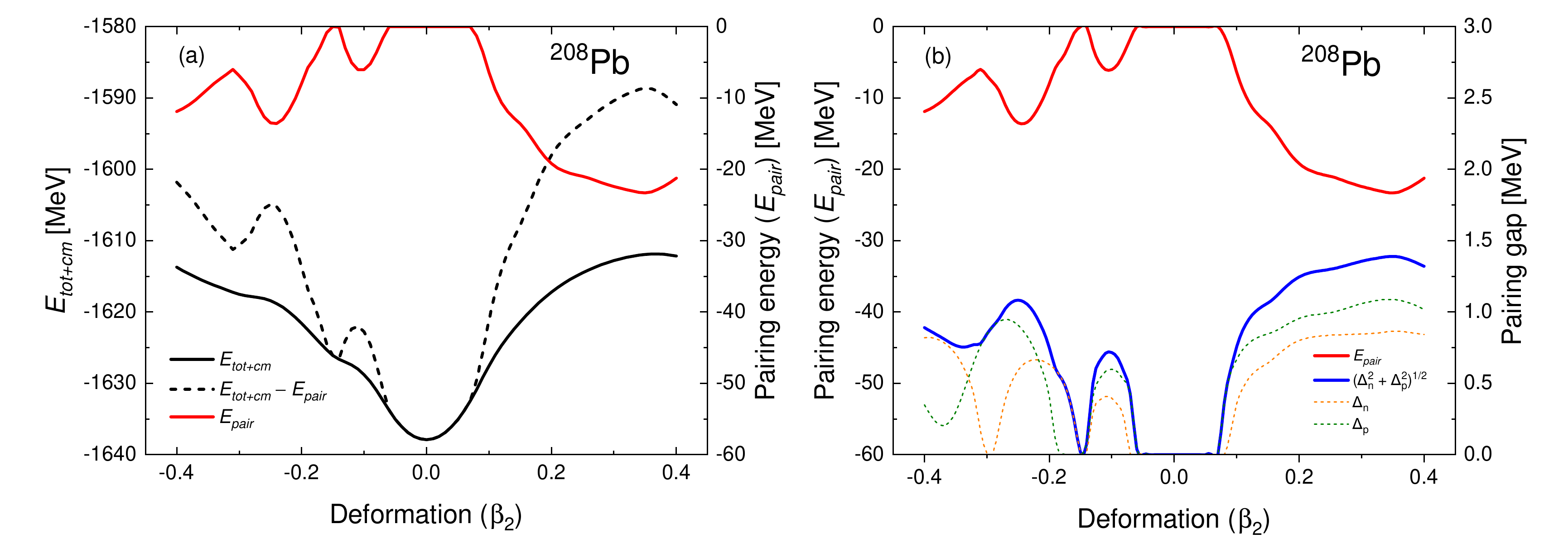}
\caption{(Color online) (Left) Evolution of the mean field energy $E_{\rm tot+cm} - E_{\rm pair} = E_{\rm EDF} + E_{\rm cm}$ (black dashed curve), pairing energy $E_{\rm pair}$ from Eq. (\ref{eq:Penergy}) (red solid curve), and their sum (black solid curve), {\it i.e.} total binding energy (TBE), $E_{\rm tot+cm}$ from Eq. (\ref{eq:EDF}), as a function of the deformation parameter $\beta_2$ for $^{208}$Pb. The pairing gaps are the accumulated integral sums of the pairing gap obtained by Eq. (\ref{eq:gap}). Main intervals in the vertical axis for both mean field energy (left y-axis) and pairing energy (right y-axis) are the same (10 MeV). (Right) {Evolution of the pairing energy and gaps as a function of the deformation parameter $\beta_2$}.   
The red solid curve shows the pairing energy, while the pairing gap of neutron ($\Delta_n$) and proton ($\Delta_p$) calculated by Eq. (\ref{eq:gap1}), and the average of the two pairing gaps, $\sqrt{\Delta_n^2 + \Delta_p^2}$, are shown by red dashed, green dashed and blue solid curves, respectively.  
Note that the axis scale of the pairing gap is smaller by a factor of 20 than the pairing energy scale.}
\label{fig1}
\end{figure}

We calculated the total binding energy (TBE) using Eq. (\ref{eq:EDF}), and subtracted the pairing energy obtained by Eq. (\ref{eq:Penergy}) (See Appendix A). 

Figure~\ref{fig1} displays the evolution of the mean field energy $E_{tot+cm} - E_{pair}$, pairing energy $E_{pair}$, and their sum, {\it i.e.} TBE $E_{tot+cm}$, for $^{208}$Pb. All results are obtained by the DRHBc theory, which explicitly includes the center of mass corrections.
The TBE and the mean field energy show a local minimum at the spherical point ($\beta_2 =$0). The pairing energy is also zero at the $\beta_2 =$ 0 point, but increases negatively with the deformation as shown by the red curve in the panel (a) and (b). 

The panel (b) shows the evolution of total pairing gaps, and of that of neutrons and protons, respectively. The pairing gaps are calculated by the $V^2_k$- (or $U_k V_k$-) weighted average of the pairing gap from Eq. (\ref{eq:gap}) for neutron and proton single-particle-state in a given pairing window limited by $E_{cut}^{q.p.}$ = 100 MeV \cite{Bender2000}, 
\begin{equation}\label{eq:gap1}
 \Delta_{n,p} = {  {\sum_k \int d^3 {\bm r} V_k^2 ( {\bm r} ) \Delta ({\bm r}) } \over {\sum_k \int d^3 {\bm r} V_k^2 ( {\bm r} )  } }.
\end{equation}
The pairing gaps and the pairing energies show almost {anti-symmetric} evolution pattern. 


This can be understood simply by using a schematic solution of the gap equation \cite{Ring1980,Nilsson1995} for heavy nuclei: $E_{pair} \sim -\frac{1}{2}\overline{\rho}({\bigtriangleup_n^2} + {\bigtriangleup_p^2}) $, where $\overline{\rho}$ is the level density at the Fermi energy.
The pairing energy should be negatively correlated with $\Delta$, although both $\Delta$ and $\bar\rho$ depend on $\beta_2$ in non-trivial ways.

An interesting point is that both pairing gaps of neutrons and protons are zero at the spherical deformation, but deviate from zero with the increase of the deformation. 
This can be understood by the following argument. For any clear energy minimum which is associated with closed shells, and the level density is low around the Fermi surface,  this low level density will create small pairing gaps. On the other hand, the level density will increase away from the closed-shell, spherical energy minimum and consequently the pairing gap will become larger. In this way, 
the cross-talk nature will be established between the deformation minimum and the pairing gap through the level densities of energy minimum in the self-consistent microscopic calculations. 

\begin{figure}
\centering
\includegraphics[width=0.90\linewidth]{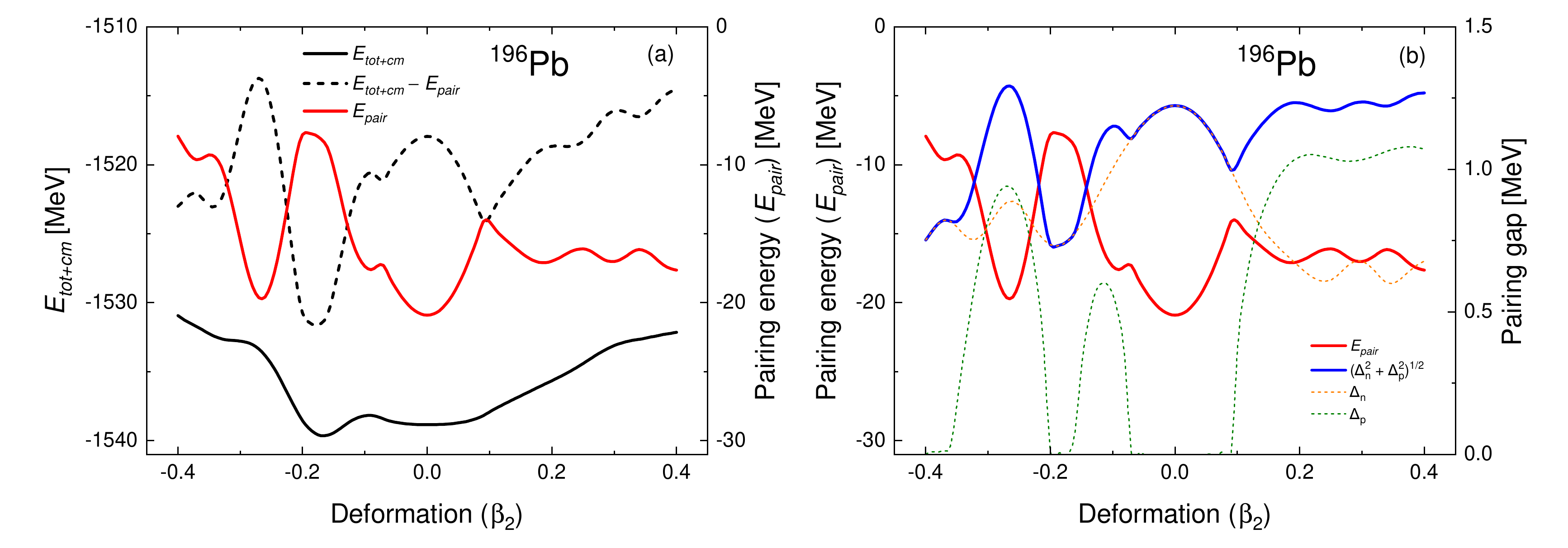}
\includegraphics[width=0.90\linewidth]{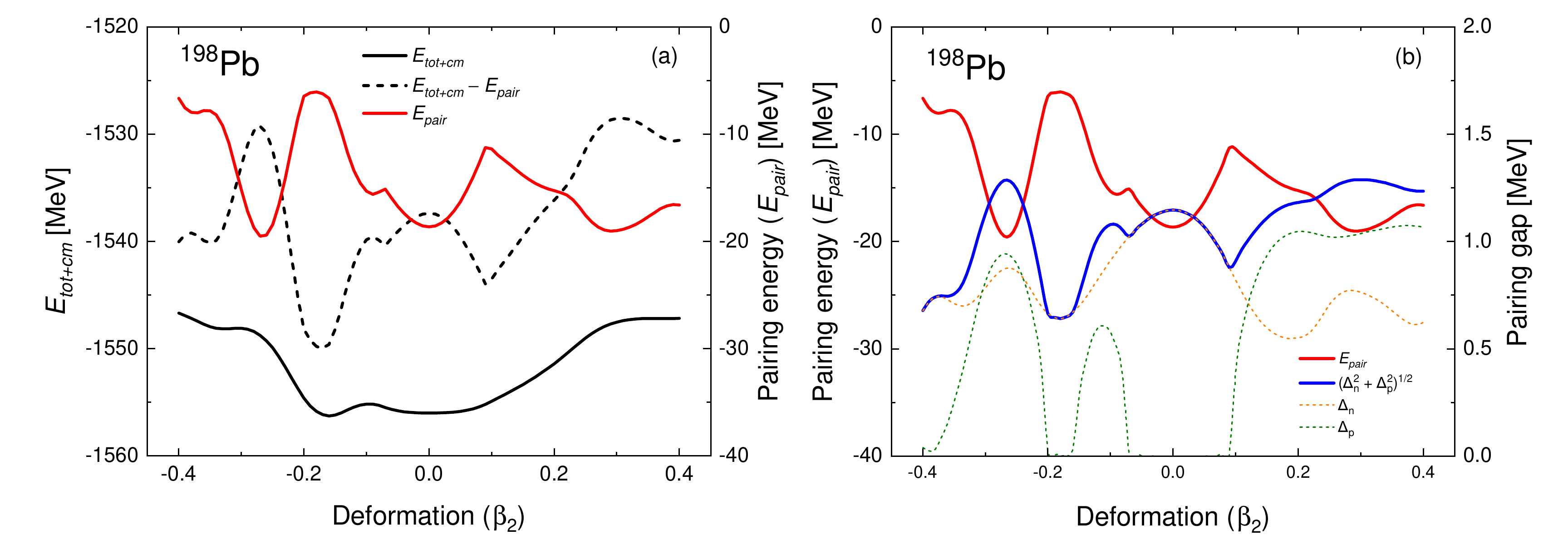}
\caption{(Color online) Same as Fig. \ref{fig1}, but for oblate deformed $^{196,198}$Pb isotopes.}
\label{fig2}
\end{figure}

Figure \ref{fig2} shows the results for oblate deformed Pb isotopes, $^{196,198}$Pb. Similarly to the spherical $^{208}$Pb, the evolution of the pairing energy is anti-correlated just to that of mean field energy. Namely, with the more binding is produced by the mean field energy, the smaller pairing energy is obtained. 
When the mean-field energy is minimum, even in a deformed configuration, it means that there is a local energy gap around the Fermi energy between
occupied and unoccupied states. Then, the level density is ``locally'' lower and the pairing gap is also lower.
By the compensation of both energies, the TBE seems to be relatively insensitive to the deformation compared to the pairing energy and the mean field energy.

The evolution of the pairing energies in the panels (b) quite resembles that of pairing gaps with a minus sign as was noticed before, although the scale in the pairing gap is different by a factor of 20 in all panels.  The small deviation in the asymmetric evolution comes from {level density differences at each deformation}. Therefore, even for oblate deformed nuclei, the evolution of mean field energy is quite similar to that of the pairing gap like the in case of spherical nucleus, despite the energy scale difference.

\begin{figure}
\centering
\includegraphics[width=0.90\linewidth]{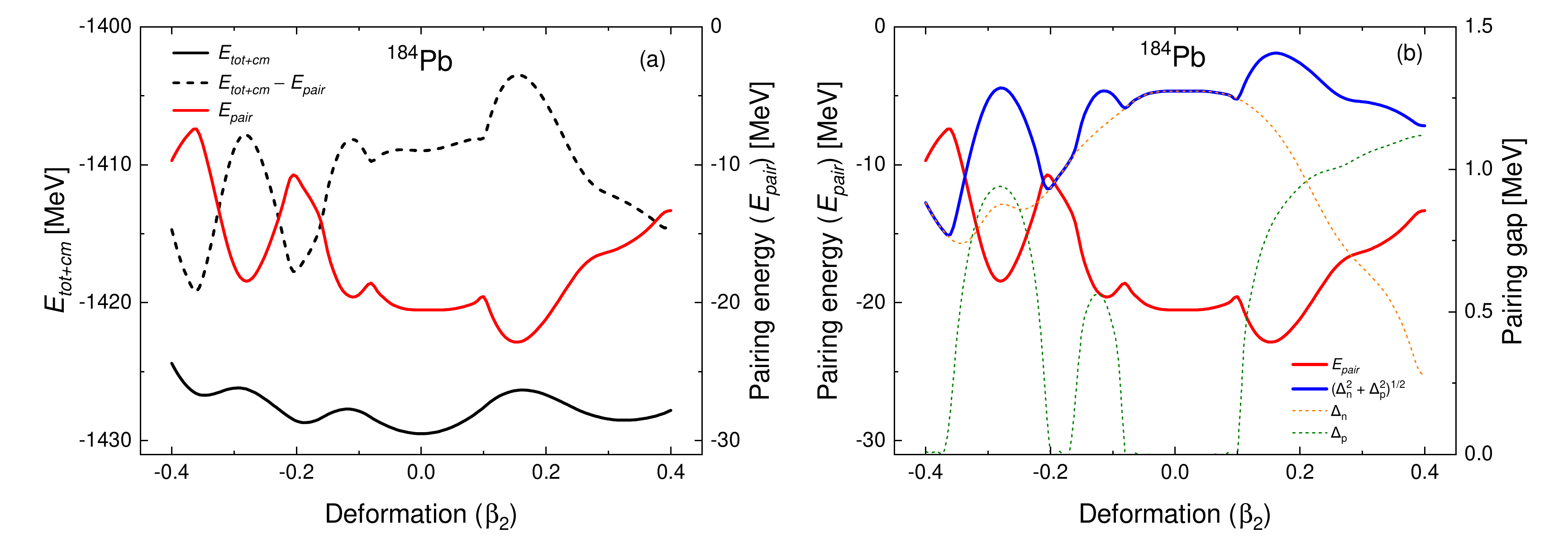}
\includegraphics[width=0.90\linewidth]{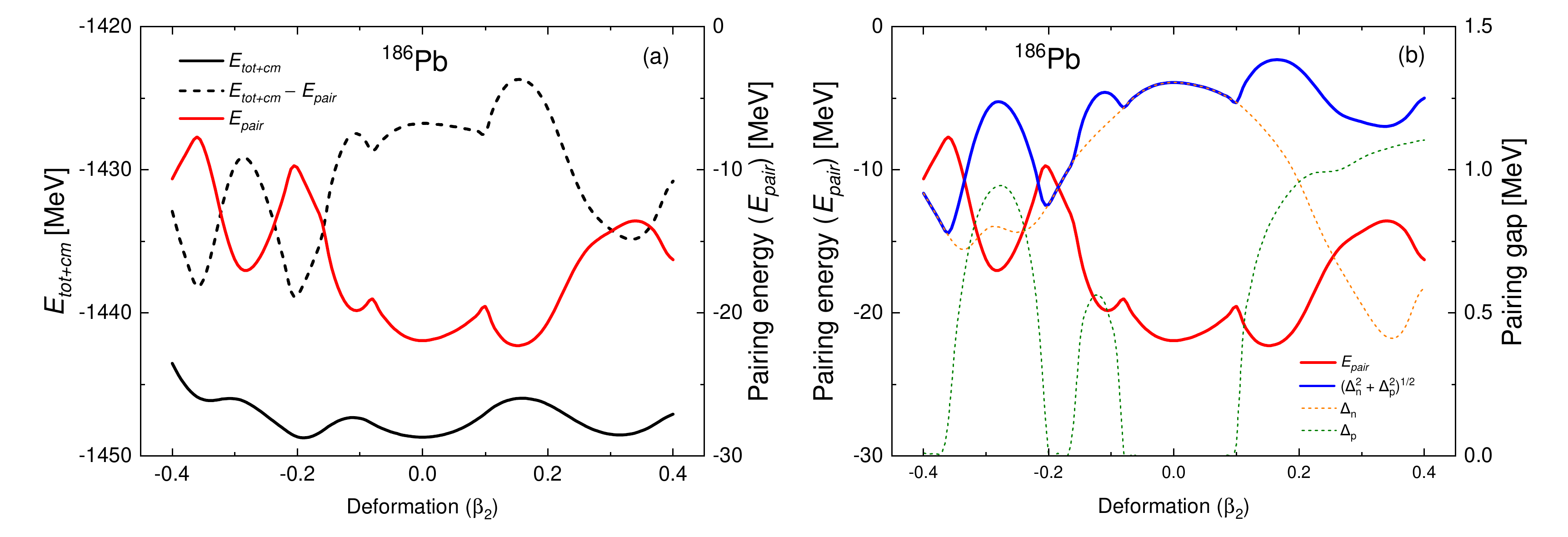}
\caption{(Color online) Same as Fig. \ref{fig1}, but for nuclei that are possible candidates of shape coexistence, namely the $^{184,186}$Pb isotopes.}
\label{fig3}
\end{figure}

In Fig. \ref{fig3}, we provide the results for nuclei that are {possible candidates of shape coexistence}, $^{184,186}$Pb \cite{Kim2022}. A similar evolution pattern to the results in the spherical and oblate cases, in Figs. \ref{fig1} and \ref{fig2}, is  found for the mean field energy, pairing energy and pairing gap. We note {the similarity of the black dashed curve in the panel (a) with the blue solid curve in the panel (b).}  It means again that with a smaller binding (in terms of mean field energy), we get a stronger pairing energy and a larger pairing gap. Contrarily, with a stronger binding, a weaker pairing energy and a smaller pairing gap appear even in soft nuclei showing the possibility of shape coexistence. 

\begin{figure}
\centering
\includegraphics[width=0.90\linewidth]{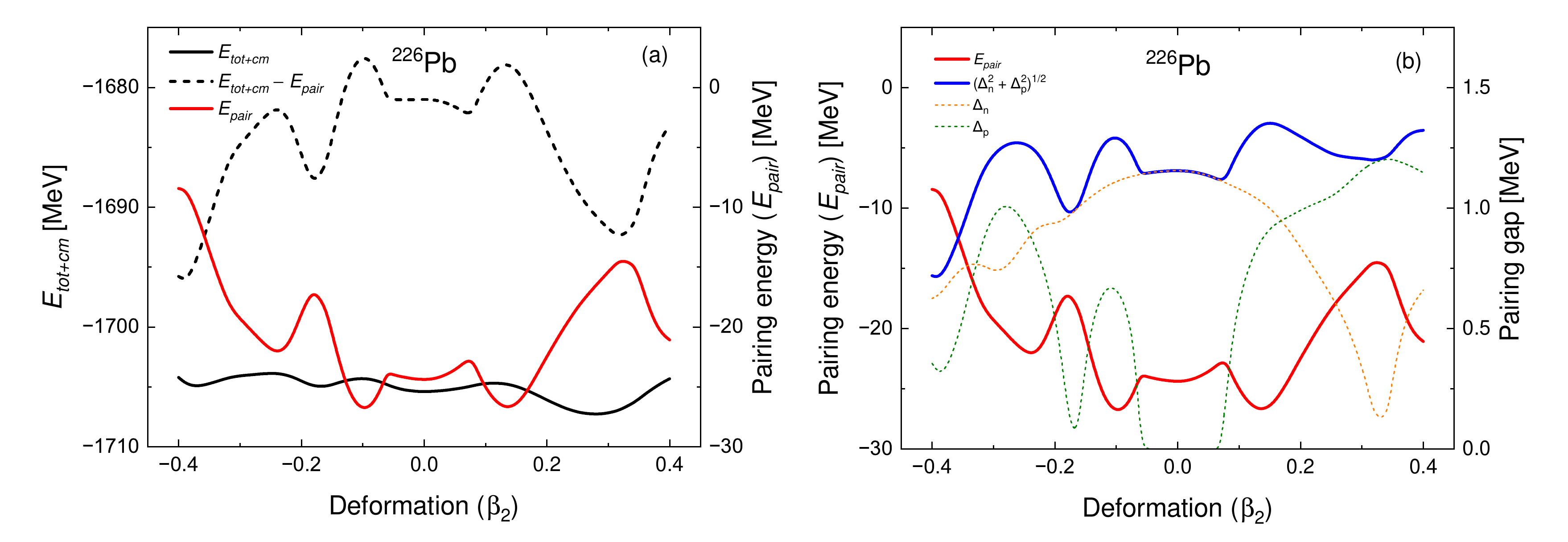}
\caption{(Color online) Same as Fig. \ref{fig1}, but for the prolate deformed $^{226}$Pb isotope.}
\label{fig4}
\end{figure}


\begin{figure}
\centering
\includegraphics[width=0.90\linewidth]{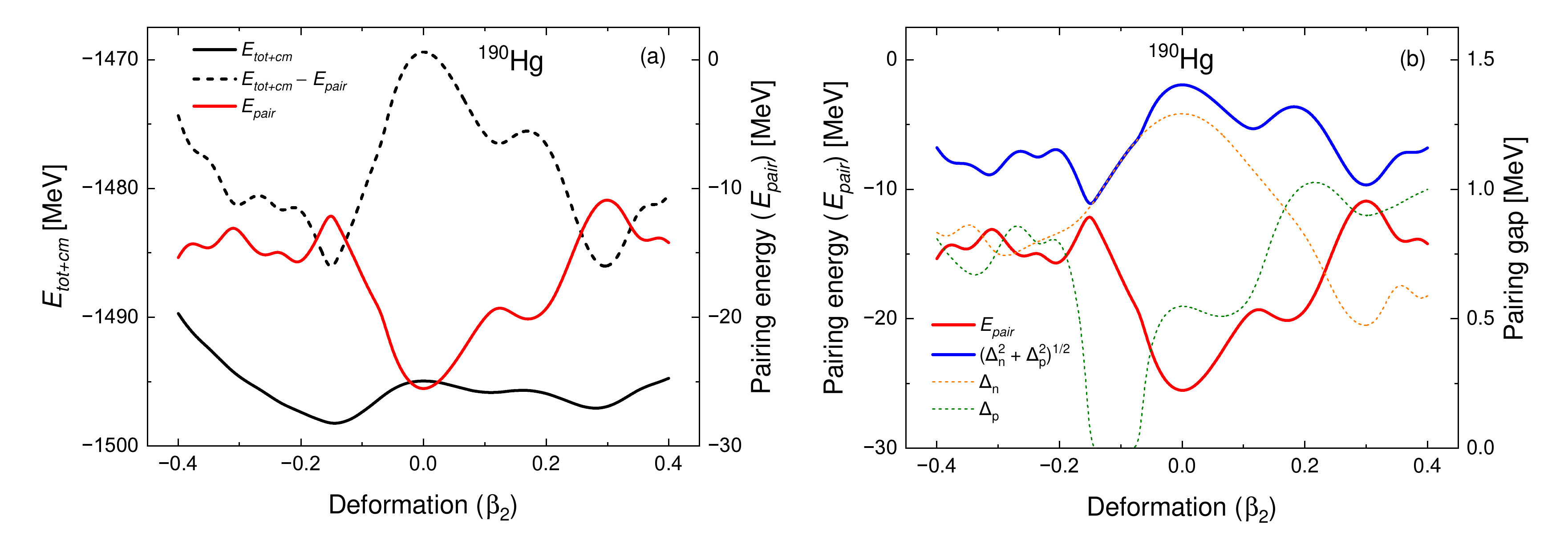}
\includegraphics[width=0.90\linewidth]{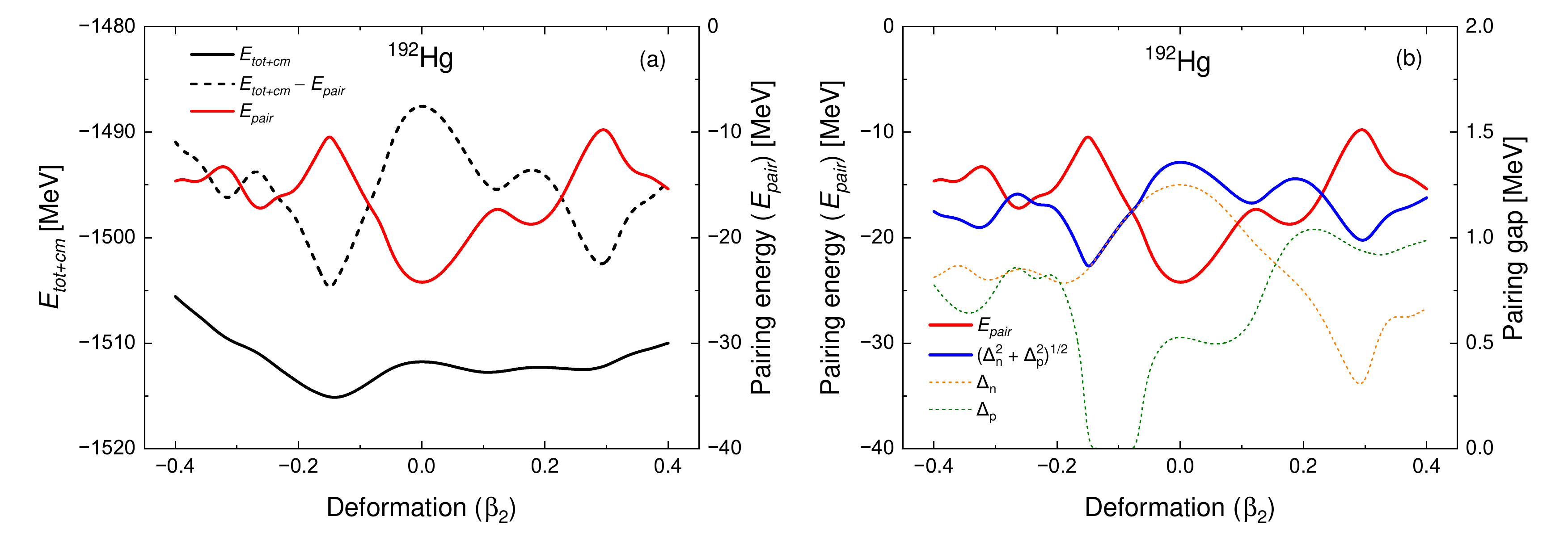}
\caption{(Color online) Same as Fig. \ref{fig1}, but for oblate deformed $^{190,192}$Hg isotopes.}
\label{fig5}
\end{figure}
In particular, $^{186}$Pb is a well-known shape coexistence nucleus \cite{Andreyev2000}, which exhibits a triplet of different nuclear shapes. 
Furthermore, recent data for the band structures of $^{186}$Pb provide detailed level schemes of the band structures for the oblate and prolate deformations \cite{Ojala2022}. Since the present calculations present the detailed pairing energy for each minimum in prolate and oblate deformed regions, we can discuss, in principle, the moment of inertia for each case. The lower panel (b) in Fig. \ref{fig3} shows larger pairing energy in the prolate region (see the absolute value of the red solid curves) than in the oblate region, which means a smaller moment of inertia {due to stronger pairing} and, consequently, a larger energy spacing in the rotational band structure. This consequence of our prediction is quite consistent with the experimental data \cite{Ojala2022}.

For prolate nuclei, we show the results in Fig. \ref{fig4}. We definitely find an evolution pattern similar to the oblate deformed cases illustrated in the previous figures. The present results mean 
that the pairing energies and paring gaps well reflect the evolution of the mean field energy by talking to each other. Another interesting point is the change of $\Delta_{n}$ and $\Delta_{p}$. In particular, the almost zero $\Delta_{p}$ near spherical shape increases with the deformation, but $\Delta_{n}$ shows opposite behavior. The zero proton pairing gap, $\Delta_p$, at $\beta_2 \approx$ 0, coming from the magic number $Z = 82$,  can be  finite in the strong deformation region.  Despite this feature of  $\Delta_p$, the averaged pairing gaps $\Delta = \sqrt{\Delta_{n}^2 + \Delta_{p}^2}$ follow the same pattern as the mean field energy evolution, at least in the nuclei considered in this work.

\begin{figure}
\centering
\includegraphics[width=0.90\linewidth]{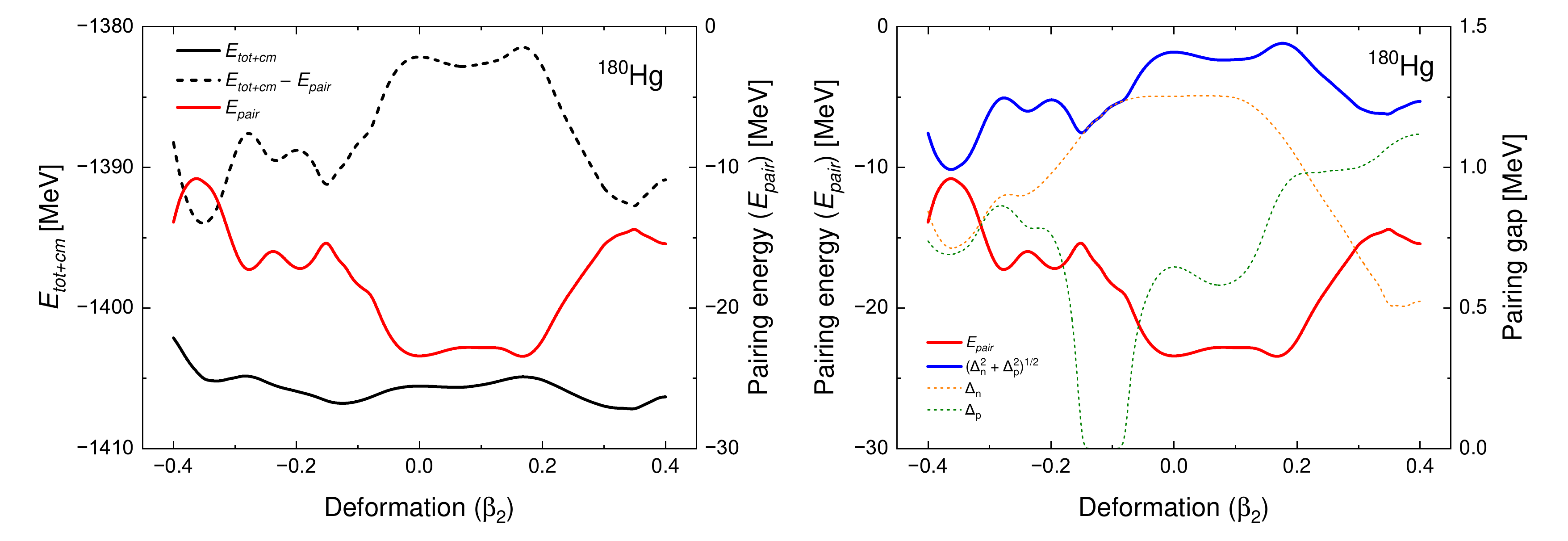}
\includegraphics[width=0.90\linewidth]{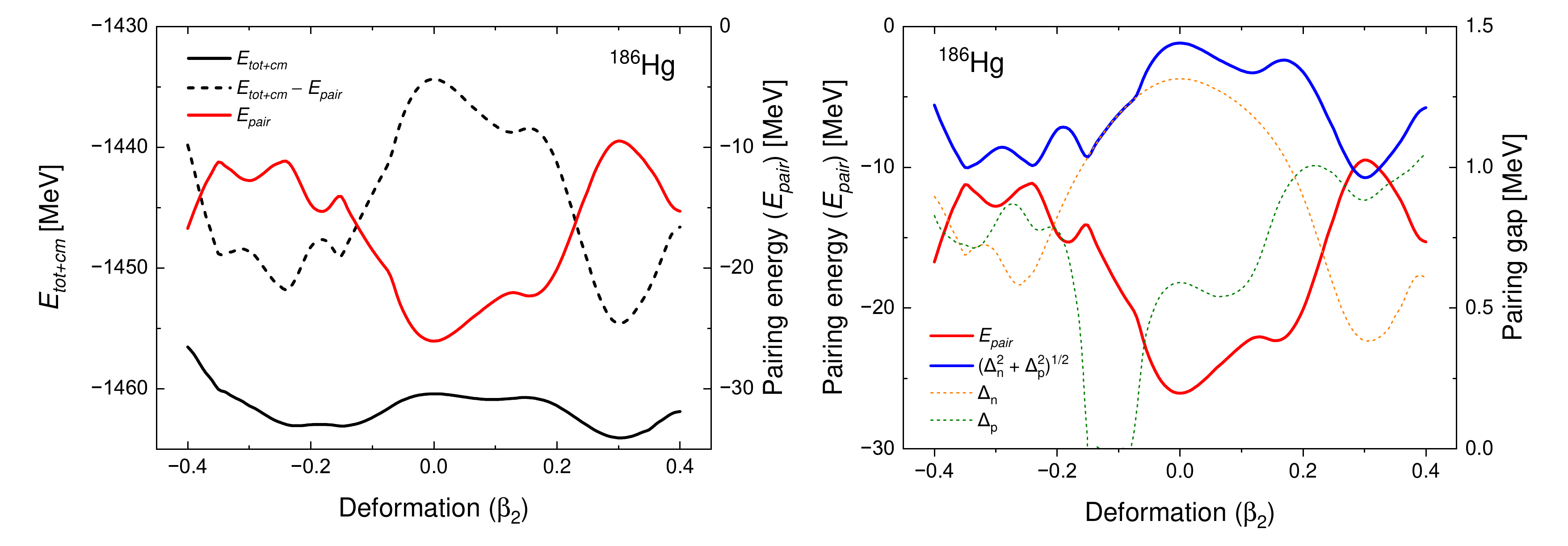}
\caption{(Color online) Same as Fig. \ref{fig1}, but for the prolate deformed $^{180,186}$Hg isotopes.}
\label{fig6}
\end{figure}

Figure \ref{fig5} and \ref{fig6} show the results for oblate and prolate Hg isotopes. Irrespective of the deformation shape, mean field energy, pairing energy and pairing gaps display almost the same patterns as in the Pb isotopes. On the other hand, the minimum points of proton pairing gaps $\Delta_p$ are a bit deviated from the spherical position because of the non-magic number.

\begin{figure}
\centering
\includegraphics[width=0.90\linewidth]{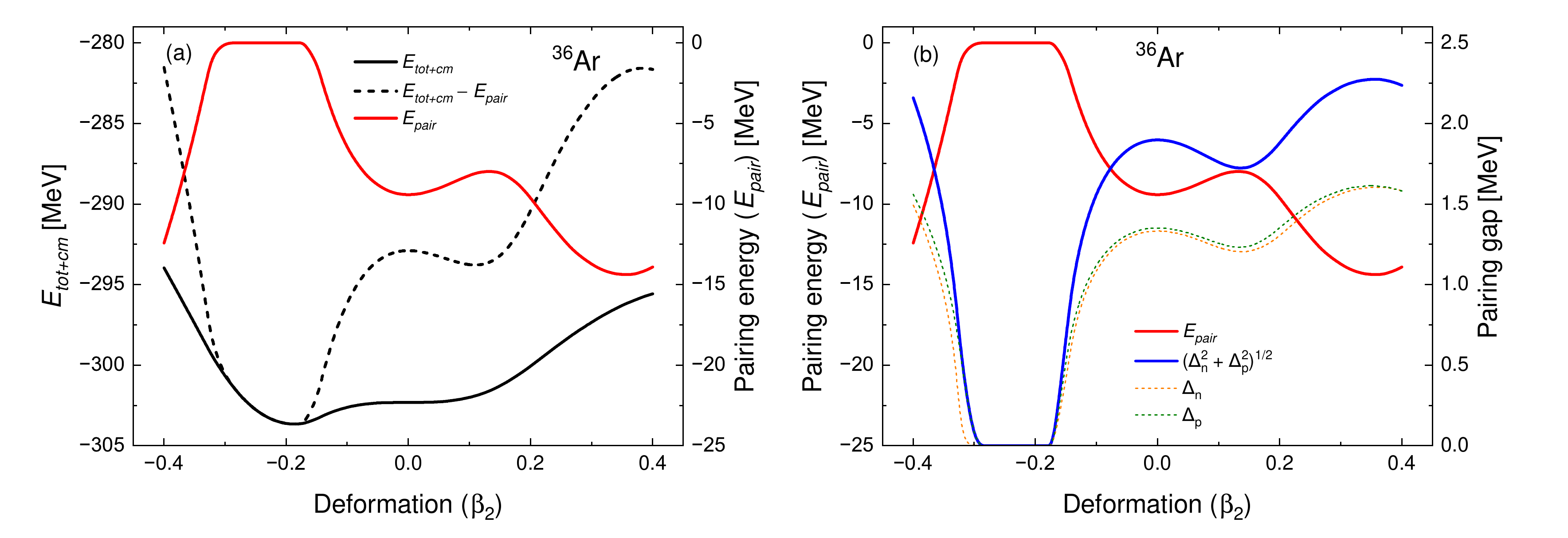}
\includegraphics[width=0.90\linewidth]{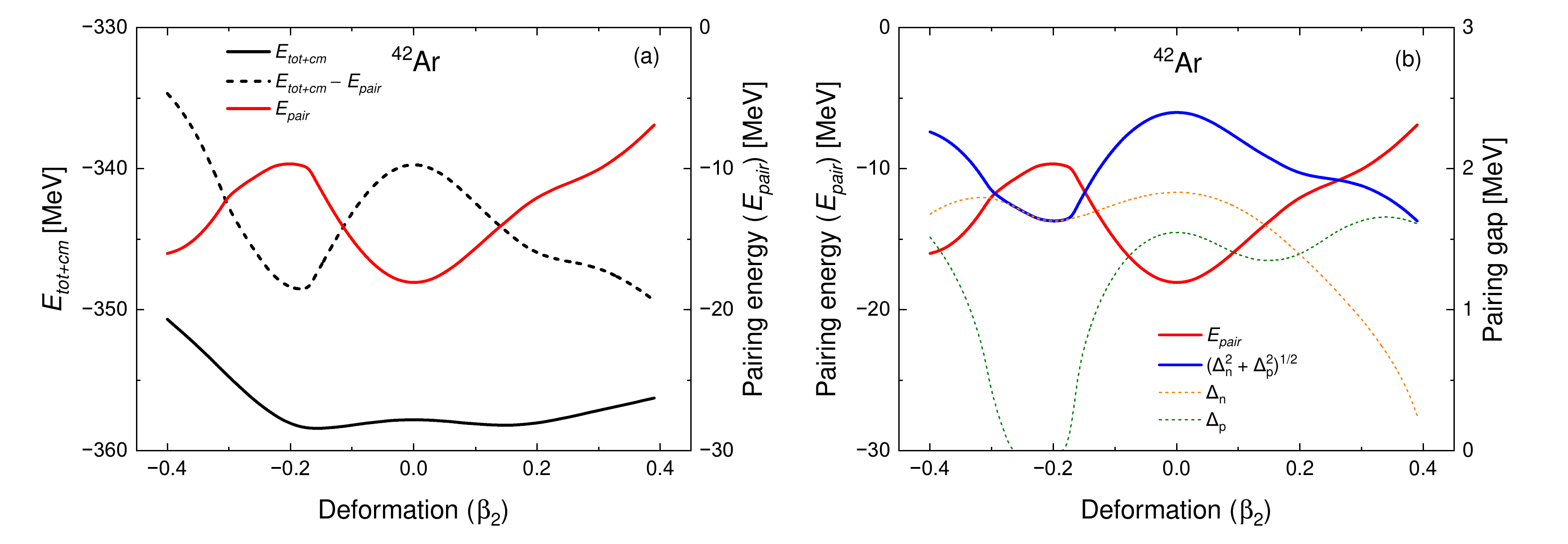}
\includegraphics[width=0.90\linewidth]{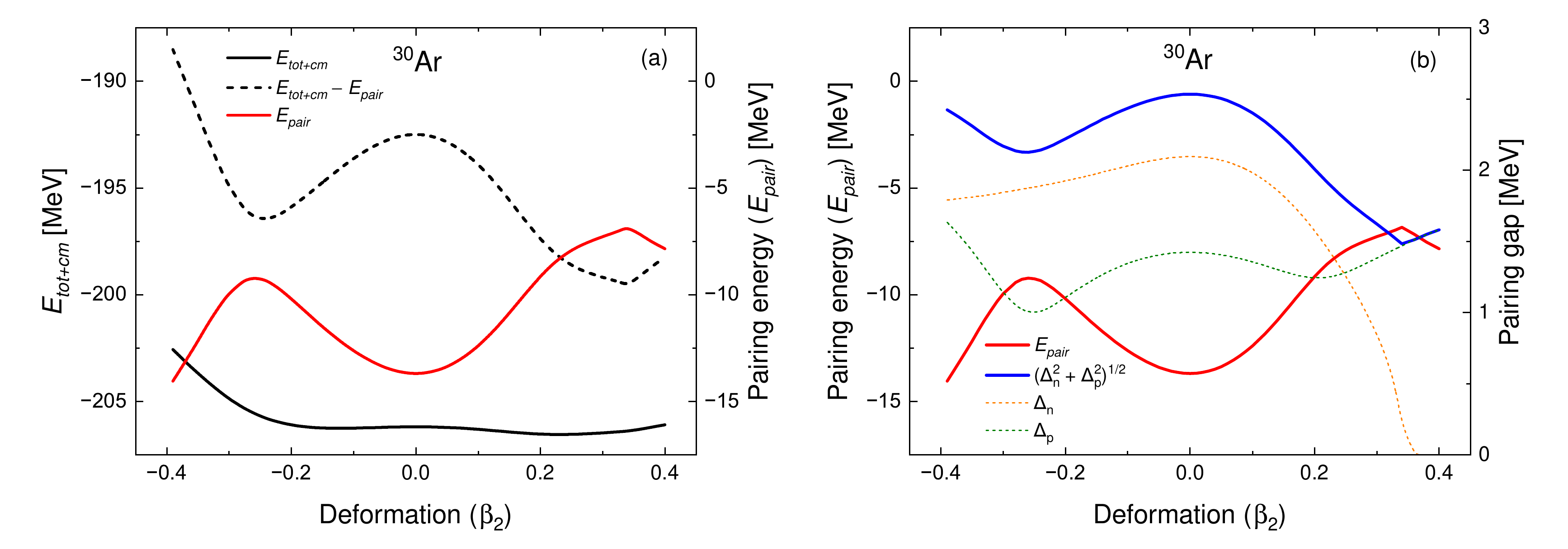}
\caption{(Color online) Same as Fig. \ref{fig1}, but for oblate deformed $^{36,42}$Ar and prolate deformed $^{30}$Ar isotopes.}
\label{fig7}
\end{figure}

The discussion so far has been done for heavy nuclei. One of the important point to notice is the evolution of pairing energy and its pairing gaps. They show almost {anti-symmetric} evolution pattern. 
Figure \ref{fig7} shows the results for oblate $^{36,42}$Ar and prolate $^{30}$Ar isotopes. One may notice similar evolutions of the pairing energy, mean field energy and the pairing gap. But a subtle deviation in the {anti-symmetric} evolutions of the pairing gap and the paring energy is found in the Ar isotope case compared to {the cases of heavy nuclei}  Hg and Pb isotopes. {This difference suggests a small change of the level density evolution in the deformed light nuclei compared to the heavy nuclei.} 

The c.m. correction energy is calculated microscopically, $E_{c.m.} = - { 1 \over {2 m A}} <{\hat {\bm P}}^2>$ with $A$ the mass number and ${\hat {\bm P}} = \sum_{i}^A {\hat{\bm p}}_i$  the total momentum in the c.m. frame. It has been shown that the microscopic c.m.
correction provides more reasonable and reliable results than the phenomenological corrections \cite{Kaiyuan2022,Bender2000}.

\subsection{Results by DSHF+BCS calculation}

In order to examine whether the phenomena calculated by the DRHBc model in Section IIIA are general, we perform non-relativistic deformed Skyrme HF+BCS (DSHF+BCS) calculations as well. The essential difference between DRHBc and DSHF+BCS are the contributions of exchange terms in the nucleon-nucleon and Coulomb interactions: the exchange terms are discarded in the relativistic model, while they are included in the non-relativistic HF model.
We adopt the SLy4 functional \cite{Chabanat1998} and a constant-gap approximation in the BCS part with $\Delta = 12/A^{1/2}$ MeV. 
Smooth cutoffs at $\pm 2.5$ MeV around the Fermi energies are applied by means of a Fermi function as was done in Ref. \cite{Krieger1990}.

\begin{figure}
\centering
\includegraphics[width=0.49\linewidth]{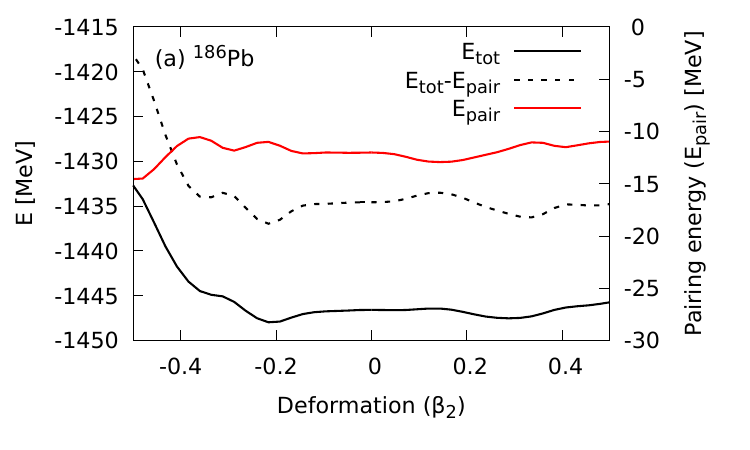}
\includegraphics[width=0.49\linewidth]{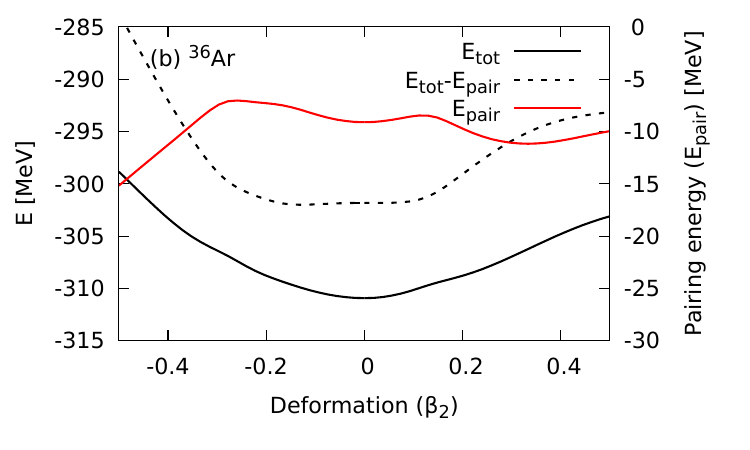}
\caption{(Color online) {Evolution of the total energy $E_{\rm tot}$, the pairing energy $E_{\rm pair}$, and the mean-field energy $E_{\rm tot}-E_{\rm pair}$ for (a) $^{186}$Pb and (b) $^{36}$Ar obtained by the DSHF+BCS calculation.} }
\label{fig8}
\end{figure}

The results in Fig. \ref{fig8} show also clear cross-talk patterns, {\it i.e.} an anti-symmetric evolution of the mean field (HF) energy and the pairing energy. Therefore, the anti-symmetric evolution is a robust phenomenon irrespective of the types of EDFs, i.e.   
even if we change the EDF for the study. 
Here we note that the oscillations of the pairing energy are much more pronounced in the DRHBc case than in the DSHF+BCS case. Also in the case of $^{36}$Ar, the DRHBc pairing energy 
is a sort of step-like function with a large drop around $\beta\approx 0.1$, while in the Skyrme case is quite mild. These quantitative differences may be an effect of the constant gap approximation. 
Nonetheless, the cross-talk we have discussed above persists.

\section{SUMMARY AND CONCLUSION}
In summary, we studied the evolution of the total binding energies and the pairing gain energies of the Pb, Hg and Ar isotopes, {as a function of the $\beta_{2}$ deformation parameter}, using the DRHBc framework. 
From a detailed analysis of the evolution of the mean field energies, pairing energies and pairing gaps, we have found that the evolution of pairing gaps, which are small and have an order of magnitude of few MeV, 
show a clear (positive) correlation to the mean field energies, which are much larger and are of the order of few hundreds MeV.  The pairing energy evolves in an anti-symmetric way with respect to the mean field energy  in all the nuclei studied in the present work. This trend, the weaker (stronger) binding energy in the mean field being associated with the stronger (weaker) paring energy, is almost irrespective of the magnitude and type of deformations. Our finding indicates that the pairing correlations are fully under cross-talk with the mean field energy.

One may argue about the model dependence on this finding. In the DRHBc approach, the model shows small deviations from experimental binding energy in neutron-deficient nuclei,  which may come from yet unknown correlations beyond the present theoretical approach, such as 
 tri-axial deformation, the exchange term in the Coulomb interaction, or Fock term contributions that are not taken into account in the PC-PK1 EDF. To confirm the universality of our finding in DRHBc model, we tested the case of $^{186}$Pb  and $^{36}$Ar by using the non-relativistic {HF+BCS approach, and found that the anti-symmetric evolution of the HF energy and the pairing energy is clearly displayed in the same way as the DRHBc calculation, 
despite quantitative differences.}

In conclusion, the nuclear shape is largely associated with  the competition of the mean field energy and the pairing energy, and both energies are evolved in anti-symmetric way as a function of the deformation. As a consequence, {the pairing gap turns out to evolve in a symmetric way with respect to the mean field energy.} 
The reason is as follows: the deformed energy minimum in nuclei takes place commonly in an area of the Nilsson-type diagram where the level density at the Fermi surface is low, {\it i.e.} {there is usually an energy gap around the Fermi surface for the shape in which} nuclei find the energy minimum or local minimum.
Consequently, this driving mechanism  to the energy minimum will be against the pairing correlation, which prefers  a high level density far from the closed shell region.  In an opposite way, the pairing correlation will be enhanced by the high level density region in which nuclei may not gain large binding energy (enough to make any energy minimum).
We can  find these cross-talks in all the  figures of the present study. This finding could help to understand quantitatively the relationship of the mean field energy and the pairing energy as a gross feature of many  nuclei in a wide region of the mass table. The study using other nuclear models are left as a future work.

\section{Acknowledgements}
Helpful discussions with members of the DRHBc Mass Table Collaboration are gratefully appreciated. {We thank also Peter Ring and Nobuo Hinohara for enlightening discussions on the deformation and pairing.} This work was supported by the National Research Foundation of Korea (Grant Nos. NRF-2020R1A2C3006177, NRF-2021R1F1A1060066, NRF-2021R1A6A1A03043957 and {RS-2024-00361003}). This work was supported by the National Supercomputing Center with supercomputing resources including technical support (KSC-2022-CRE-0333).

\appendix

\section{Derivation of HFB energy expressed in terms of quasiparticle energies}

In this appendix, we derive the expression of the total HFB energy as given in 
Eqs. \eqref{eq:EDF} and \eqref{eq:EDF2}, using the HFB equations (see also Ref. \cite{Meng1998}). 
We follow the notations of Ref. \cite{Ring1980}. 

The total HFB energy is given as
\begin{align}
E_{\rm HFB} &= \sum_{ij}t_{ij}\rho_{ji}
+ \frac{1}{2}\sum_{ijkl}\bar v_{ijkl}\rho_{ki}\rho_{lj} 
+ \frac{1}{4}\sum_{ijkl}\bar v_{ijkl}'\kappa_{ij}^*\kappa_{kl} , 
\end{align}
where $t_{ij}$ is the the matrix element of kinetic energy, 
$\bar v_{ijkl}$ ($\bar v_{ijkl}'$) is the antisymmetrized matrix element of 
the effective interaction in the ph (pp) channel, and
\begin{align}
\rho_{ij} &= \langle a_j^\dagger a_i\rangle,\ \ \ 
\kappa_{ij} = \langle a_j a_i\rangle
\end{align}
are the density matrix and the pairing tensor, respectively, 
with $a_i$ ($a_i^\dagger$) being the annihilation (creation) operator of the state $i$. 
{Note that $\rho$ is hermitian while $\kappa$ is antisymmetric.}
By introducing the HF field $\Gamma$ and the pairing field $\Delta$, defined as 
\begin{align}
\Gamma_{ij} &= \sum_{kl}\bar v_{ikjl}\rho_{lk},\ \ \ h = t+\Gamma,
\nonumber
\\
\Delta_{ij} &= \frac{1}{2}\sum_{kl}\bar v_{ijkl}'\kappa_{kl}, 
\nonumber
\end{align}
{the total HFB energy is expressed as}
\begin{align}
E_{\rm HFB} &= 
{\rm Tr}[t\rho] + \frac{1}{2}{\rm Tr}[\Gamma\rho] - \frac{1}{2}{\rm Tr}[\Delta\kappa^*]
=
{\rm Tr}[h\rho] - \frac{1}{2}{\rm Tr}[\Gamma\rho] - \frac{1}{2}{\rm Tr}[\Delta\kappa^*]
\\
&= E_{\rm HF} + E_{\rm pair}, 
\label{eq:E_HFB}
\end{align}
where $E_{\rm HF} = {\rm Tr}[t\rho] + \frac{1}{2}{\rm Tr}[\Gamma\rho] = {\rm Tr}[h\rho] - \frac{1}{2}{\rm Tr}[\Gamma\rho]$, and 
$E_{\rm pair} = - \frac{1}{2}{\rm Tr}[\Delta\kappa^*]$. 
{Notice that $\Gamma$ is hermitian and $\Delta$ is antisymmetric. }

{
The HFB equations for $U$ and $V$ {given by} the generalized Bogoliubov transformation} read
\cite{Ring1980},
\begin{align}
E_\alpha U_{i\alpha} &= 
\sum_j\left[ (h_{ij}-\delta_{ij}\lambda)U_{j\alpha} + \Delta_{ij}V_{j\alpha}\right], 
\\
E_\alpha V_{i\alpha} &= 
\sum_j\left[-\Delta_{ij}^*U_{j\alpha} - (h_{ij}^*-\delta_{ij}\lambda)V_{j\alpha} \right],
\label{eq:HFBeq} 
\end{align}
where $\alpha$ labels the quasiparticle states, $E_\alpha$ is the quasiparticle energy, 
and $\lambda$ is the Fermi energy. 
Note that $U$ and $V$ are related to $\rho$ and $\kappa$ as 
\begin{align}
\rho &= V^*V^T, \ \ \ \kappa = V^*U^T.
\end{align}
By multiplying both sides of Eq. \eqref{eq:HFBeq} by $V_{i\alpha}^*$ 
and summing over $\alpha$ and $i$, one obtains
\begin{align}
\sum_\alpha (E_\alpha-\lambda)\sum_{i}V_{i\alpha}^*V_{i\alpha}
&= \sum_\alpha
\sum_{ij}\left[ -h_{ji}V_{i\alpha}^*V_{\alpha j}^T 
- \Delta_{ij}^*V_{i\alpha}^*U_{\alpha j}^T\right]
\nonumber\\
&=
-{\rm Tr}[h\rho]+{\rm Tr}[\Delta^*\kappa], 
\end{align}
{with}
\begin{align}
{\rm Tr}[h\rho] &= 
\sum_\alpha (\lambda-E_\alpha)\sum_{i}|V_{i\alpha}|^2 - 2E_{\rm pair}. 
\label{eq:Tr(hrho)_2}
\end{align}
Inserting Eq. \eqref{eq:Tr(hrho)_2} into $E_{\rm HF}$, 
one finally finds for the HF part of the total HFB energy, 
\begin{align}
E_{\rm HF} &= {\rm Tr}[h\rho] - \frac{1}{2}{\rm Tr}[\Gamma\rho]
=
\sum_\alpha (\lambda-E_\alpha)\sum_{i}|V_{i\alpha}|^2
- \frac{1}{2}{\rm Tr}[\Gamma\rho]
- 2E_{\rm pair}. 
\end{align} 
This corresponds to Eq. \eqref{eq:EDF2}. 
Notice that, in our framework, we employ a Hartree EDF rather than a HF one, based on a density-dependent effective interaction.

\end{document}